\documentstyle[preprint,aps,epsf,graphicx]{revtex}
\begin{document}

\tightenlines

\title{Properties of  D-mesons in nuclear matter within a self-consistent 
coupled-channel approach}

\author{L. Tol\'os, J. Schaffner-Bielich, A. Mishra}
\address{Institut f\"ur Theoretische Physik 
J.W. Goethe Universit\"at,
Frankfurt am Main}

\date{\today}

\maketitle

\begin{abstract}
The spectral density of the $D$-meson in the nuclear environment is
studied within a self-consistent coupled-channel approach assuming a
separable potential for the bare meson-baryon interaction. 
The $DN$ interaction, described through a G-matrix, generates dynamically
the $\Lambda_c$ (2593) resonance. This resonance is the charm counterpart 
of the $\Lambda$ (1405) resonance generated from the s-wave $\bar{K}N$ 
interaction
in the I=0 channel. The medium modification of the D-meson 
spectral density due to the Pauli blocking of intermediate states
as well as due to the dressing of the D-mesons, nucleons and pions is investigated. We observe that the inclusion of coupled-channel effects and the self-consistent dressing of the $D$-meson results in an overall reduction of the in-medium $D$-meson changes compared to previous work which neglect those effects.

\vspace{0.5cm}



\end{abstract}


\section{Introduction}
\label{sec:intro}

The study of the properties of hadrons in a hot and dense medium is
an important problem in strong interaction physics. It is a topic of active
research interest as it has direct implications in heavy-ion collision
experiments, as well as in the study of astrophysical compact objects like
neutron stars. The medium modifications of hadrons are probed in the
relativistic heavy-ion collision experiments. In particular, at CERN-SPS,
the experimentally observed dilepton spectra \cite{ceres,helios} 
have been attributed to 
the medium modifications of the spectral properties of the vector mesons,
especially of the $\rho$-meson \cite{Brat1,CB99,vecmass,dilepton,liko}
and can not be explained by the vacuum
hadronic properties. Furthermore, the $K^{\pm}$ production from nuclear collisions
at GSI-SIS energies of 1-2 AGeV have shown that in-medium properties of the kaons
have been seen in the collective flow pattern of $K^+$ mesons as well as
in the abundancy and spectra of antikaons
\cite{CB99,cmko,lix,Li2001,K5,K6,K4,kaosnew,cassing-laura}. The medium modifications 
of $D$($\bar D$) mesons, which show analogy to the $\bar K$($K$) mesons
resulting from replacing s-quark (s-antiquark) by the c-quark (c-antiquark),
have also become a subject of recent interest
\cite{arata,liuko,friman,weise,digal,qmc}.
The medium changes for $D$-mesons can have important
consequences for the open charm enhancement in nucleus-nucleus
collisions \cite{cassing} as well as for J/$\Psi$ suppression
as observed at the SPS \cite{NA501}. The NA50
Collaboration has claimed to see an open charm enhancement by up to a
factor of three in central $Pb+Pb$ collisions at 158 A$\cdot$GeV
\cite{NA50e}.
The medium modifications of the open charm mesons ($D$($\bar D$))
can also modify the high mass ($M > 2\;\; GeV$) dilepton spectra \cite{lin}
since they can be produced abundantly in high energy heavy-ion collisions.
Transverse momentum spectra of electrons from $Au+Au$ collisions at $\sqrt{s}=130$ GeV have been measured at midrapidity by the PHENIX experiment at RHIC \cite{phenix}. The spectra show an excess above the background from decays of light hadrons and photon conversion. The observed signal is consistent with that expected from semi-leptonic decays of charmed mesons.

In high
energy heavy-ion collisions at RHIC ($\sqrt s \sim$ 200 GeV), an appreciable
contribution of J/$\Psi$ suppression is expected to be due to the
formation of a quark-gluon Plasma (QGP) \cite{blaiz}.  However, the medium modification of $D$-mesons should modify the $J/\Psi$ absorption 
in  hot and dense nuclear medium and 
can provide a possible explanation for $J/\Psi$ suppression.
 The
effect of the hadron absorption of J/$\Psi$'s is found to be not negligible
\cite{zhang,brat5,elena}. It is thus of importance to understand the
interactions of the $D$-mesons in the hadronic medium for 
$J/\Psi$ production as well.

At finite densities, the medium modification of the $D$-meson mass 
has been studied using the QCD sum rule (QSR) approach \cite{arata,weise}.
Due to the presence of a light quark in the D-meson, the mass modification of
the $D$-meson has a large contribution from the light quark condensates.
The large shift of mass ($\simeq$ 50 MeV at  nuclear matter density,
$\rho_0$) in the $D$-mesons is originated from the contribution of the 
$m_c \langle \bar q q \rangle_N$ term in the operator product expansion. 
The quark-meson coupling (QMC) model
predicts a mass drop of the $D$-meson to be of the order of 60 MeV at
$\rho=\rho_0$ \cite{qmc}, which is very similar to the value
obtained in the QCD sum rule calculations of Refs.~\cite{arata,weise}.
At finite temperatures, the studies of quarkonium dissociation
\cite{digal,wong} using heavy quark potential from lattice QCD \cite{lattice}
suggest a similar drop of the $D$-meson mass.

In a recent work, the mass modification of the $D$-meson 
in hot and dense matter, 
arising due to its interaction with the light hadron sector,
was studied in a chiral effective model 
\cite{dmeson}. The chiral SU(3) model, used to study the hadronic properties
in the hot hyperonic matter \cite{kristof1} was generalized to 
SU(4) to include the charmed mesons. Interactions
of the $D$-meson with the light hadron sector were derived to study 
the in-medium mass of the $D$-meson.

In all these investigations, the spectral features of a $D$-meson embedded in hot and dense matter have not been studied. Coupled-channel effects as well as dressing of intermediate propagators have been completely ignored, which turned out to be crucial for describing the strange counter part of the $D$-meson, the $\bar K$ meson in the nuclear medium.
In the present investigation, the spectral density
for the $D$-meson is calculated using a self-consistent coupled-channel 
G-matrix 
calculation. The medium effects like the Pauli blocking on the nucleons 
in the intermediate states and the effects from the dressing 
of the $D$-mesons, nucleons as well as pions are investigated. The coupled-channel formalism generates dynamically the $\Lambda_c(2593)$ resonance, like the $\Lambda(1405)$ resonance for the case of antikaons.

We have organized the present paper as follows. In
Sect.~\ref{sec:formalism} we first review our formalism. 
Our results are
presented and discussed in Sect.~\ref{sec:results} and the
concluding remarks are given in Sect.~\ref{sec:conclusions}.

\section{FORMALISM}
\label{sec:formalism}

In this section we present the formalism to obtain
the self-energy or single-particle potential
of a D-meson embedded in infinite symmetric nuclear matter.
This self-energy accounts for the interaction of the D-meson
with the nucleons and
its calculation requires the knowledge of the in-medium $D N$
interaction, which  will be described by a $G$-matrix.
The medium effects incorporated in this $G$-matrix include
the Pauli blocking on the nucleons in the intermediate states as well as 
the dressing of the D-meson, nucleons and pions.

\subsection{In-medium s-wave $D N$ interaction}

\begin{figure}[htb]
\centerline{
     \includegraphics[width=0.6\textwidth]{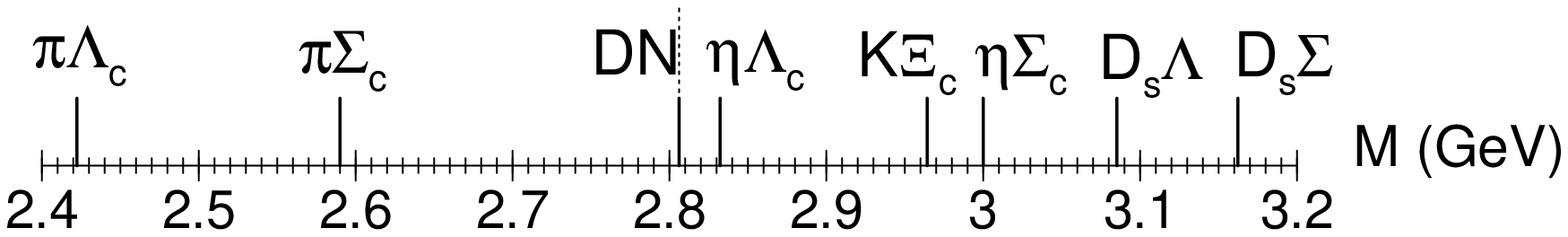}
}
      \caption{\small Mass thresholds
}
        \label{fig:dmeson0}
\end{figure}

The effective $D N$ interaction in the nuclear medium or
$G$-matrix is obtained from the bare $D N$ interaction, in a similiar way as done in Ref.~\cite{Tolos01} for the $\bar{K} N$ interaction. However, in this case, only the in-medium s-wave $D N$ interaction is considered taking, as a bare interaction, a separable potential model as done for the $K^- p$ interaction in Ref.~\cite{Koch}.
For the separable potential we use the following ansatz in momentum space
\begin{eqnarray}
V_{i,j}(k,k')&=&g^2 C_{i,j} v_i(k) v_j(k')\nonumber\\
&=&\frac{g^2}{\Lambda^2}C_{i,j}\Theta(\Lambda-k)\Theta(\Lambda-k') \ ,  
\end{eqnarray}
where $g$ is the coupling constant and $\Lambda$ the cutoff. These
two parameters will be determined by fixing the position and the width of the $\Lambda_c(2593)$ resonance, the analogous 
 to the $\Lambda(1405)$ resonance in the charm sector.
For the interaction matrix $C_{ij}$, we use the standard result derived from SU(3) flavor symmetry (~\cite{Koch,Weise98}). 
This bare interaction allows for the transition from $D N$ to other channels, namely, $\pi \Lambda_c$, $\pi \Sigma_c$, $\eta \Lambda_c$ and $\eta \Sigma_c$, all having charm $c=1$. Therefore, we are confronted with  a coupled-channel problem.
In principle, one should use SU(4) symmetry as we incorporate charmed mesons. However, channels with a strange and charm quarks  such as  $K \Xi_c$, $D_s \Lambda$ or $D_s \Sigma$ have not been considered due to the fact that their respective masses are well above the $D N$ threshold (see Fig.~\ref{fig:dmeson0}). Therefore, we only consider channels with up, down and charm-quark content keeping the SU(3) symmetry.

The resultant meson-baryon $G$-matrices can be grouped in a matrix
notation, where each box corresponds to one channel.
The $D N$ channel can have isospin $I=0$ or $I=1$. In the first
case, it couples  to the $\pi \Sigma_c$ and $\eta \Lambda_c$ channels and the
corresponding
matrix has the following structure

\[\left(\begin{array}{ccc}
G_{{D}N\rightarrow{D}N} &
G_{{\pi}{\Sigma_c}\rightarrow{D}N} &
G_{{\eta}{\Lambda_c}\rightarrow{D}N} \\
G_{{D}N\rightarrow{\pi}{\Sigma_c}} &
G_{{\pi}{\Sigma_c}\rightarrow{\pi}{\Sigma_c}} &
G_{{\eta}{\Lambda_c}\rightarrow{\pi}{\Sigma_c}} \\
G_{{D}N\rightarrow{\eta}{\Lambda_c}}  &
G_{{\pi}{\Sigma_c}\rightarrow{\eta}{\Lambda_c}}  &
G_{{\eta}{\Lambda_c}\rightarrow{\eta}{\Lambda_c}}
        \end{array}
  \right) \ ,\]

while for $I=1$ it can couple to  the $\pi \Lambda_c$, $\pi \Sigma_c$  and $\eta \Sigma_c$ channels

\[\left(\begin{array}{cccc}
G_{{D}N\rightarrow{D}N} &
G_{{\pi}{\Lambda_c}\rightarrow{D}N} &
G_{{\pi}{\Sigma_c}\rightarrow{D}N} &
G_{{\eta}{\Sigma_c}\rightarrow{D}N} \\
G_{{D}N\rightarrow{\pi}{\Lambda_c}}  &
G_{{\pi}{\Lambda_c}\rightarrow{\pi}{\Lambda_c}} &
G_{{\pi}{\Sigma_c}\rightarrow{\pi}{\Lambda_c}} &
G_{{\eta}{\Sigma_c}\rightarrow{\pi}{\Lambda_c}} \\
G_{{D}N\rightarrow{\pi}{\Sigma_c}} &
G_{{\pi}{\Lambda_c}\rightarrow{\pi}{\Sigma_c}} &
G_{{\pi}{\Sigma_c}\rightarrow{\pi}{\Sigma_c}} &
G_{{\eta}{\Sigma_c}\rightarrow{\pi}{\Sigma_c}} \\
G_{{D}N\rightarrow{\eta}{\Sigma_c}}  &
G_{{\pi}{\Lambda_c}\rightarrow{\eta}{\Sigma_c}}&
G_{{\pi}{\Sigma_c}\rightarrow{\eta}{\Sigma_c}} &
G_{{\eta}{\Sigma_c}\rightarrow{\eta}{\Sigma_c}}
        \end{array}
  \right) \ .\]

Keeping this structure in mind, the $G$-matrix is formally given by

\begin{eqnarray}
\langle M_1 B_1 \mid G(\Omega) \mid M_2 B_2 \rangle &&= \langle M_1 B_1
\mid V \mid M_2 B_2 \rangle   \nonumber \\
&& \hspace*{-2cm}+\sum_{M_3 B_3} \langle M_1 B_1 \mid V \mid
M_3 B_3 \rangle
\frac {Q_{M_3 B_3}}{\Omega-E_{M_3} -E_{B_3}+i\eta} \langle M_3 B_3 \mid
G(\Omega)
\mid M_2 B_2 \rangle \ .
   \label{eq:gmat1}
\end{eqnarray}

In Eq.~(\ref{eq:gmat1}),  $M_i$ and $B_i$  represent the possible
mesons ($D$, $\pi$, $\eta$) and
baryons ($N$, $\Lambda_c$, $\Sigma_c$) respectively, and their corresponding
quantum numbers such as spin, isospin, charm, and linear momentum.
The function $Q_{M_3,B_3}$ stands for the Pauli operator which allows
only intermediate nucleon states
compatible with the Pauli principle. The energy variable $\Omega$ is
the
so-called starting energy  while $\sqrt{s}$ is the invariant
center-of-mass energy, i.e., 
$\sqrt{s}=\sqrt{\Omega^2-P^2}$, where $(\Omega,\vec{P}\,)$ is
the total meson-baryon four momentum in a frame in which nuclear matter is
at rest.

The former equation for the $G$-matrix has to be considered together with
a prescription for the
single-particle energies of all the mesons and baryons participating
in the reaction and in the intermediate states. These energies can be
written as
\begin{equation}
 E_{M_i(B_i)}(k)=\sqrt{k^2 +m_{M_i(B_i)}^2} + U_{M_i(B_i)}
(k,E_{M_i(B_i)}^{qp}) \ ,
\label{eq:spen}
\end{equation}
where $U_{M_i(B_i)}$ is the  single-particle
potential of each meson (baryon) calculated at the real
quasiparticle energy $E_{M_i(B_i)}^{qp}$. For baryons, this quasiparticle-energy is given by
\begin{equation}
E_{B_i}^{qp}(k)=\sqrt{k^2 +m_{B_i}^2} + {\mathrm {Re}\,} 
U_{B_i}(k,E_{B_i}^{qp}) \ ,
\label{eq:qp}
\end{equation}
while, for mesons, it is obtained by solving the following equation
\begin{equation}
(E_{M_i}^{qp}(k))^2=k^2 +m_{M_i}^2+{\mathrm {Re}\,}\Pi_{M_i}(k,E_{M_i}^{qp}) \ .
\end{equation}

In the present paper we have considered the single-particle potential for the $D$-meson, 
nucleons and pions together with the decay width of the $\Sigma_c$ meson. 
The reason 
for the inclusion of this width will be clarified in the results section.

For nucleons, as done in Ref.~\cite{Tolos02},
we have used a relativistic $\sigma-\omega$ model,
where the scalar and vector coupling constants, $g_s$ and $g_v$
respectively, are density dependent \cite{Mach89}. 

One of the most important modifications comes from the introduction of the
pion self-energy, $\Pi_{\pi}(k,\omega)$, in the intermediate $\pi\Lambda_c$
and $\pi\Sigma_c$ states present in the construction of
the effective $DN$  interaction. The pion is dressed with the
momentum and energy-dependent self-energy \cite{pion}
 which was also studied in Ref.~\cite{Tolos02}. This self-energy 
incorporates a p-wave piece
built up from the coupling to $1p-1h$, $1\Delta-1h$ and $2p-2h$
excitations, together with short-range correlations. The model also
contains a small and constant s-wave part.

The $D$-meson  single-particle potential
in the Brueckner-Hartree-Fock approach is
schematically given by
\begin{equation}
 U_{D}(k,E_{D}^{qp})= \sum_{N \leq F} \langle D N \mid
 G_{D N\rightarrow
D N} (\Omega = E^{qp}_N+E^{qp}_{D}) \mid D N \rangle,
\label{eq:self}
\end{equation}
where the summation over nucleon states is limited by the nucleon Fermi
momentum. 
As it can be easily seen from Eq.~(\ref{eq:self}), since the
effective $DN$  interaction ($G$-matrix) depends on the
$D$-meson single-particle energy, which in turn depends on the
$D$-meson potential, we are confronted
with a self-consistent problem.
We also note that the
$G$-matrix in the above equation becomes complex
due to the possibility of $D$$N$ state decaying into the $\pi\Lambda_c$ and $\pi\Sigma_c$
channels.
As a consequence, the potential
$U_{D}$ is also a complex quantity. Further details are given in the next section.

The $G$-matrix equation for the s-wave in-medium $DN$ interaction in the partial wave basis
 using the quantum numbers of the
relative and
center-of-mass motion reads
\begin{eqnarray}
\langle (M_1B_1);k''|
      G^{I}(P,\Omega) | (M_2B_2);k  \rangle
     &&=  
     \langle (M_1B_1);k'' |
      V^{I} | (M_2B_2);k \rangle 
      \nonumber \\
    &&\hspace*{-5cm} +\sum_{M_3 B_3}
      \int \frac{k'^{2}}{2\pi^2}dk'
      \langle (M_1 B_1);k'' |
      V^{I} | (M_3 B_3); k'\rangle
\nonumber
\\
      && \hspace*{-4.5cm}\times \frac{\overline{Q}_{M_3 B_3}(k',P)}{\Omega
-\sqrt{m_{B_{3}}^2+\widetilde{k_{B_{3}}^2}}                             
-
\sqrt{m_{M_{3}}^2+\widetilde{k_{M_{3}}^2}}
-U_{B_{3}}(\widetilde{k_{B_{3}}^2})-U_{M_{3}}(\widetilde{k_{M_{3}}^2})
+i\eta}
\nonumber \\
      && \hspace{-4.5cm}\times \langle (M_3 B_3);k' |
      G^{I}(P,\Omega) | (M_2B_2);k \rangle \ ,
   \label{eq:gmat}
\end{eqnarray}
where the variables $k$, $k'$, $k''$ 
are the relative momenta and $P$ is the linear
center-of-mass momentum.
The functions $\widetilde{k_{B}^2}$ and $\widetilde{k_{M}^2}$ are,
respectively, the square of the momentum of the baryon and that of the
meson in the intermediate states, averaged
over the angle between the total momentum ${\vec P}$ and the relative
momentum $\vec{k}'$ (see appendix A in Ref.~\cite{Tolos01})
\begin{eqnarray}
\widetilde{k_B^2}(k',P)&=&k'^2 + \left(\frac{m_B}{m_{
M}+m_B}\right)^2
P^2 \nonumber \ ,\\
\widetilde{k_{M}^2}(k',P)&=&k'^2 + \left(\frac{m_{M}}{m_{
M}+m_B}\right)^2  
P^2 \ .
\end{eqnarray}
The angle average of the Pauli operator, $\overline{Q}_{M_3 B_3}(k',P)$,
differs from unity only for the ${D}N$ channel (for details see appendix A in 
Ref.~\cite{Tolos01}).

\subsection{$D$-meson single-particle energy in the
Brueckner-Hartree-Fock approximation}

For the s-wave component of the $D N$ interaction, the Brueckner-Hartree-Fock 
approximation to the single-particle potential
of a $D$-meson embedded in a Fermi sea of nucleons 
[Eq.~(\ref{eq:self})] becomes
\begin{eqnarray}
 U_{D}(k_{D},E_{D}^{qp}(k_{D})) &=
& \frac{1}{2 \pi^2}
\sum_{I}(2I+1) (1+\xi)^3 \int_{0}^{k_{max}} \, k^2dk \,
f(k,k_{D}) \nonumber \\
& & \times  \langle (DN) ; k | G^{I}
(\overline{P^2},E^{qp}_{D}(k_{D})+ 
E^{qp}_{N}(\overline{k_{N}^2}))
|
(D N); k \rangle  \ ,
\label{eq:upot1}
\end{eqnarray}
where $\overline{P^2}$ and $\overline{k_{N}^2}$ are the square 
of the center-of-mass momentum and nucleon momentum,
respectively, averaged
over the angle between the external ${D}$-meson momentum in the lab
system, $\vec{k}_{D}$, and the ${D}N$
relative momentum, ${\vec k}$, used as integration variable in Eq.~(\ref{eq:upot1})
 (see appendix B in Ref.~\cite{Tolos01}). These angle averages
eliminate the angular dependence
of the $G$-matrix and allow to perform the angular integration in Eq.
(\ref{eq:upot1}) analytically, giving rise to the weight function,
$f(k,k_{D})$,
\begin{equation}
f(k,k_{D})= \left\{ \begin{array}{cl} 1 & {\rm for\ }  k\leq
\frac{k_{F}-\xi k_{D}}{1+\xi} , \\ 0 &
{\rm for\ }
|\xi k_{D}-(1+\xi)k| > k_{F} , \\
\displaystyle\frac{{k_{F}}^2-[\xi k_{D}-(1+\xi)k]^2}{4\xi
(1+\xi)k_{D}k
} & \mbox{otherwise,}
\end{array} \right. \nonumber
\end{equation}
where $\xi=\displaystyle\frac{m_{N}}{m_{D}}$ and $k_F$ is the
Fermi momentum.
The magnitude of
the relative momentum 
$k$ is constrained by
\begin{equation}
   k_{max} = \frac{k_{F}+\xi k_{D}}{1+\xi} \ .
\nonumber
\end{equation}

After self-consistency for the on-shell value
$U_{D}(k_{D},E_{D}^{qp})$ is
achieved, one can obtain the complete energy dependence of the
self-energy $\Pi_D(k_D,\omega)$,
\begin{equation}
\Pi_D(k_D,\omega)=2\sqrt{k_D^2+m_D^2} \, U_{D}(k_D,\omega) \,
\label{eq:relation}
\end{equation}
by replacing $E^{qp}_{D}$ in
Eq.~(\ref{eq:upot1}) by $\omega$. This self-energy can then be used to
determine the
 $D$-meson single-particle propagator in the medium,
\begin{equation}
D_{D}(k_{D},\omega) = \frac {1}{\omega^2 -k_{D}^2 -m_{D}^2
-2 \sqrt{m_{D}^2+k_D^2} U_{D}(k_{D},\omega)} \ ,
\label{eq:prop}
\end{equation}
and the corresponding spectral density
\begin{equation}
S_{D}(k_{D},\omega) = - \frac {1}{\pi} {\mathrm Im\,} D_{D}(k_{D},\omega)\ .
\label{eq:spec}
\end{equation}

In fact, only the value of the potential 
$U_{D}$ at the quasiparticle energy $\omega=E^{qp}_{D}$ is
determined self-consistently. This amounts
to taking, in the subsequent iterations leading to
self-consistency, the so-called
``quasiparticle" approximation to the ${D}$-meson propagator
\begin{equation}
D^{qp}_{D}(k_{D},\omega) = \frac {1}{\omega^2 -k_{D}^2
-m_{D}^2
-2 \sqrt{m_{D}^2+k_D^2} U_{D}(k_{D},E^{qp}_{D})} \ ,
\label{eq:quasip}
\end{equation}
which gives rise to a simplified spectral strength
\begin{eqnarray}
&&S^{qp}_{D}(k_{D},\omega) = \nonumber \\
&&-\frac{1}{\pi}\frac {2
\sqrt{m_{D}^2+k_D^2} {\rm Im\,} U_{D}(k_{D},E^{qp}_{D})}{\mid \omega^2
-k_{D}^2 -m_{D}^2
-2 \sqrt{m_{D}^2+k_D^2} {\rm Re\,} U_{D}(k_{D},E^{qp}_{D})\mid^2 +
\mid 2 \sqrt{m_{D}^2+k_D^2} {\rm Im\,} U_{D}(k_{D},E^{qp}_{D}) \mid^2}
\ .
\label{eq:sqp}
\end{eqnarray}
The location and
width of the peak in this distribution are
determined, respectively, by the real and imaginary parts of
$U_{D}(k_{D},E^{qp}_{D})$.

This self-consistent scheme was also used in previous works for the $\bar{K}$ meson \cite{Tolos01,Tolos02} and,  
although it represents a simplification with respect to the more sophiscated scheme followed  in Refs.~\cite{Ram00,Lutz98a} for the $\bar K$ case where the full energy dependence of the $\bar K$ self-energy is self-consistent determined, the approximation is sufficiently good as already shown in Refs.~\cite{Tolos01,Tolos02}.

\section{Results}
\label{sec:results}

\begin{figure}[htb]
\centerline{
     \includegraphics[width=0.6\textwidth]{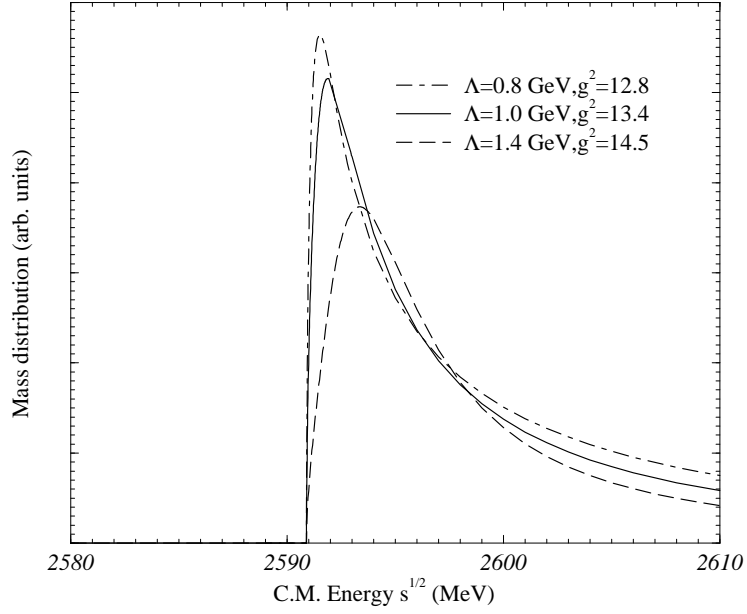}
}
      \caption{\small $\Lambda_c(2593)$ mass spectrum for different sets of coupling constants and cutoffs.
}
        \label{fig:dmeson1}
\end{figure}

We start this section by showing the mass distribution of the $\pi \Sigma_c$ state in Fig.~\ref{fig:dmeson1}.
 The mass distribution is given by
\begin{equation}
\frac{d\sigma}{dm}=C \mid T_{\pi \Sigma_c \rightarrow \pi \Sigma_c}^{I=0} \mid^2
p_{CM} 
\end{equation}
where C is related to the particular reaction generating the $\pi \Sigma_c$ state prior to final state interactions, $p_{CM}$ is the $\pi \Sigma_c$ relative momentum and 
$T_{\pi \Sigma_c \rightarrow \pi \Sigma_c}^{I=0}$ is the isospin zero component of the on-shell s-wave T-matrix for the $\pi \Sigma_c$ channel.
The study of the mass distribution of the $\pi \Sigma_c$ state in the $I=0$ channel  reflects the $\Lambda_c(2593)$ resonance as seen in 
Fig.~\ref{fig:dmeson1}. In this figure the mass distribution of the $\pi \Sigma_c$ state for $I=0$ is displayed as a function of the C.M. energy for different sets of  coupling constants $g$ and cutoffs $\Lambda$. We observe that our coupled-channel calculation generates dynamically the $\Lambda_c(2593)$ resonance. The position ($2593.9 \pm 2$ MeV) and width ($\Gamma=3.6^{+2.0}_{-1.3}$ MeV) are obtained for a given set of coupling constants and cutoffs in the range between 0.8 and 1.4 GeV. This resonance was also obtained in the framework of the $\chi$-BS(3) approach based on the chiral SU(3) Lagrangian and formulated in terms of the Bethe-Salpeter equation \cite{Lutz04}.
It is interesting to observe that the set of coupling constants and cuttoffs that generate the resonance are very similar to the ones used in the $s=-1$ sector to reproduce dynamically the $\Lambda(1405)$ resonance \cite{Koch}.

\begin{figure}[htb]
\centerline{
     \includegraphics[width=0.6\textwidth]{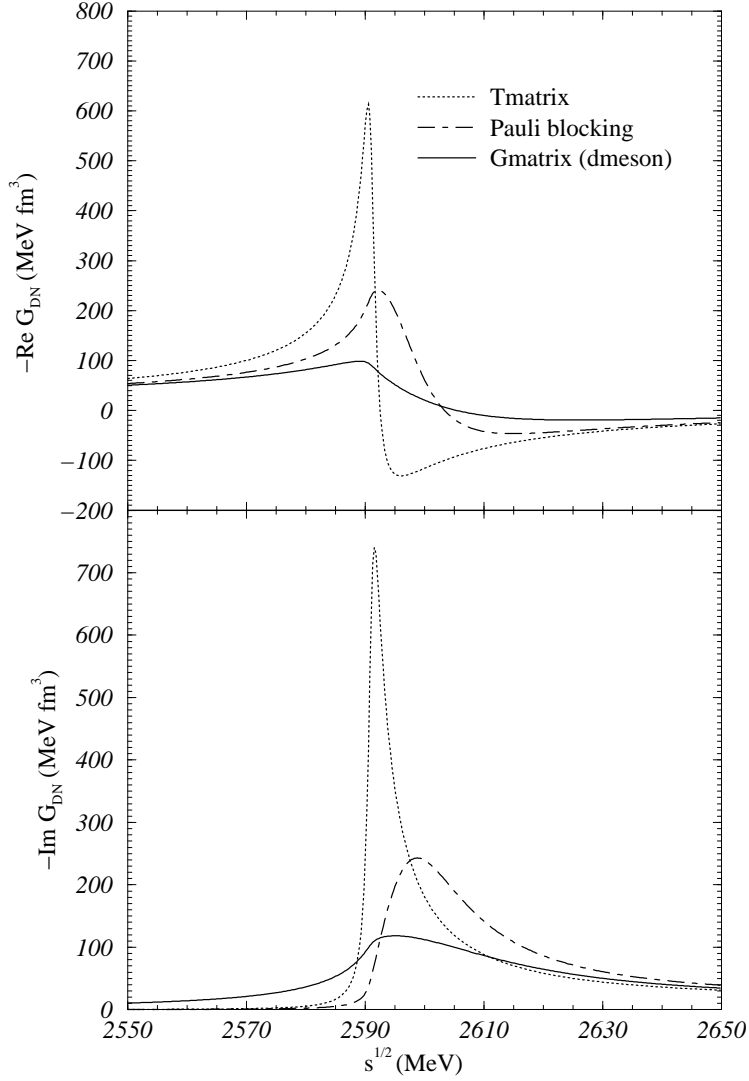}
}
      \caption{\small Real and imaginary parts of the s-wave $DN$ amplitude in the $I=0$  channel as functions of the center-of-mass energy at total momentum $\mid\vec{k}_D+\vec{k}_N\mid=0$ for $\Lambda=1$ GeV and $g^2=$13.4 and for different approaches: T-matrix calculation (dotted lines), in-medium calculation including only Pauli blocking at $\rho=\rho_0$ (dot-dashed lines) and self-consistent calculation for the $D$-meson at $\rho=\rho_0$ (solid lines).
}
        \label{fig:dmeson2}
\end{figure}

Once the position and width of the $\Lambda_c(2593)$ resonance are reproduced dynamically, we study the effect of the different medium modifications on the resonance. The real and imaginary parts of the resulting in-medium s-wave $D N$ amplitudes in the $I=0$ channel for $\Lambda=1.0$ GeV and $g^2$=13.4 and for a total momentum $\mid\vec{k}_D+\vec{k}_N\mid=0$ are given in Fig.~\ref{fig:dmeson2} as a function of the invariant center-of-mass energy for different approaches: T-matrix calculation (dotted lines), including Pauli blocking (dot-dashed lines) and self-consistent calculation for the $D$-meson (solid lines) at nuclear matter saturation density $\rho_0=0.17$ fm$^{-3}$. Due to the fact that the resonance lies just few MeV above the $\pi \Sigma_c$ threshold, 
we have included the decay width of the $\Sigma_c$ baryon in our calculations by considering the corresponding imaginary part for the $\Sigma_c$ self-energy. In this way, the resonance is reproduced in a more realistic way, even below the $\pi \Sigma_c$ state. We clearly see, as noticed already for the $\Lambda(1405)$ resonance, the repulsive effect of the Pauli blocking on the resonance, being generated at higher energies. However, as already pointed out for the $\bar K N$ interaction, the shift in energy of the $\Lambda_c(2593)$ resonance is intimately connected with the energy dependence of the $D N$ interaction and changes are expected from a self-consistent incorporation of the $D$-meson properties in the $DN$ interaction. Therefore, the effect of dressing the $D$-meson with the complete complex self-energy on the in-medium $D N$ amplitude makes the resonance peak  stay pretty close or even at lower energies than its free space location. This is due to the attraction felt by the $D$-meson that compensates the repulsive effect induced by Pauli blocking on the nucleon, as already seen for the $\Lambda(1405)$ resonance in Refs.~\cite{Tolos01,Koch,Ram00,Lutz98a,Waas}. On the other hand, the real and imaginary parts of the $D N$ amplitude become much smoother because of the $D$-meson strength being spread out over energies.

\begin{figure}[htb]
\centerline{
     \includegraphics[width=0.6\textwidth]{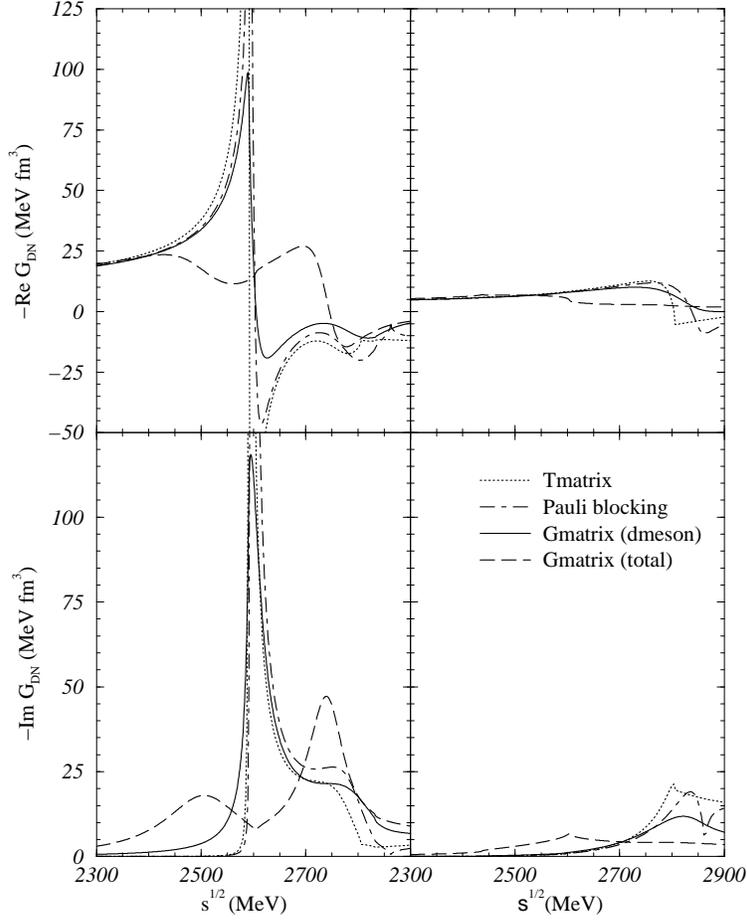}
}
      \caption{\small Real and imaginary parts of the s-wave $DN$ amplitude in the $I=0$ channel (left panels) and $I=1$  channel (right panels) as functions of the center-of-mass energy at total momentum $\mid\vec{k}_D+\vec{k}_N\mid=0$ in a larger scale of energy for $\Lambda=1$ GeV and $g^2=$13.4 and for different approaches: T-matrix calculation (dotted lines), in-medium calculation including only Pauli blocking at $\rho=\rho_0$ (dot-dashed lines), self-consistent calculation for the $D$-meson at $\rho=\rho_0$  (solid lines) and self-consistent calculation for the $D$-meson including the dressing of nucleons and the pion self-energy at $\rho=\rho_0$ (long-dashed lines).
}
        \label{fig:dmeson3}
\end{figure}

In Fig.~\ref{fig:dmeson3} we display the previous plot of  the real and imaginary parts of the s-wave $I=0$ $DN$ amplitude together with the $I=1$ channel  for a larger energy scale and a smaller  range in the y-axis. Apart from the previous three approaches, T-matrix calculation (dotted lines), in-medium calculation only including Pauli blocking effects (dot-dashed lines) and self-consistent calculation for the $D$- meson potential (solid lines), we also include the in-medium properties of the nucleons together with the pion self-energy in the self-consistent process for the D-meson potential (long-dashed lines). While the nucleon dressing only shifts the $\Lambda_c(2593)$ resonance in energy, as seen for the $\Lambda(1405)$ resonance, the pion dressing could introduce important changes to the previous self-consistent procedure. In this  approach and for $I=0$, the imaginary part of the $D N$ interaction becomes smoother in the region of energies where the $\Lambda_c(2593)$ resonance was generated in the previous three approaches. For energies below the $\pi \Sigma_c$ threshold, we observe a bump  around 2.5 GeV that,  in our self-consistent many-body approach, can decay to states such as $\pi(ph)\Sigma_c$ or $D (\Lambda_ch\pi)N$, where in parentheses we have denoted an example for the component of the $\pi$ and $D$-meson that  show up at energies below $\pi \Sigma_c$.
On the other hand, it is also interesting to observe a second structure in the $I=0$ amplitude, already seen for the three initial approaches, which lies below the $DN$ threshold of 2.806 GeV. This feature in the imaginary part around 2.75 GeV seems to be enhanced in the full self-consistent procedure. This structure, like the previous one, is also a state with the $\Lambda_c$-like quantum numbers and, in this case, it can decay to the $\pi \Sigma_c$ state. 
Whether the first resonant structure  is the in-medium $\Lambda_c(2593)$ resonance 
and the second bump is a new resonance is something that deserves further studies.  
Note that the energy region of interest for the calculation of the $D$-meson  potential for $I=0$ lies  on the right-hand side of the free $\Lambda_c(2593)$ resonance where this second structure shows up for all approaches. Actually, this is a different feature of the $D N$ interaction compared to the case of the $\bar K N$ interaction. The $\bar K N$ interaction was basically determined by the behaviour of the s-wave $I=$0 $\Lambda(1405)$ resonance in the medium. 
For the $I=1$ component of the $DN$ interaction, the self-consistent calculation tends to dilute any structure present in the T-matrix. It is seen that for the full self-consistent calculation, the $I=1$ component is considerably reduced.
A more detailed analysis concerning the $D N$ interaction and, hence, the $D$-meson potential is presented in the following plots.

\begin{figure}[htb]
\centerline{
     \includegraphics[width=0.6\textwidth]{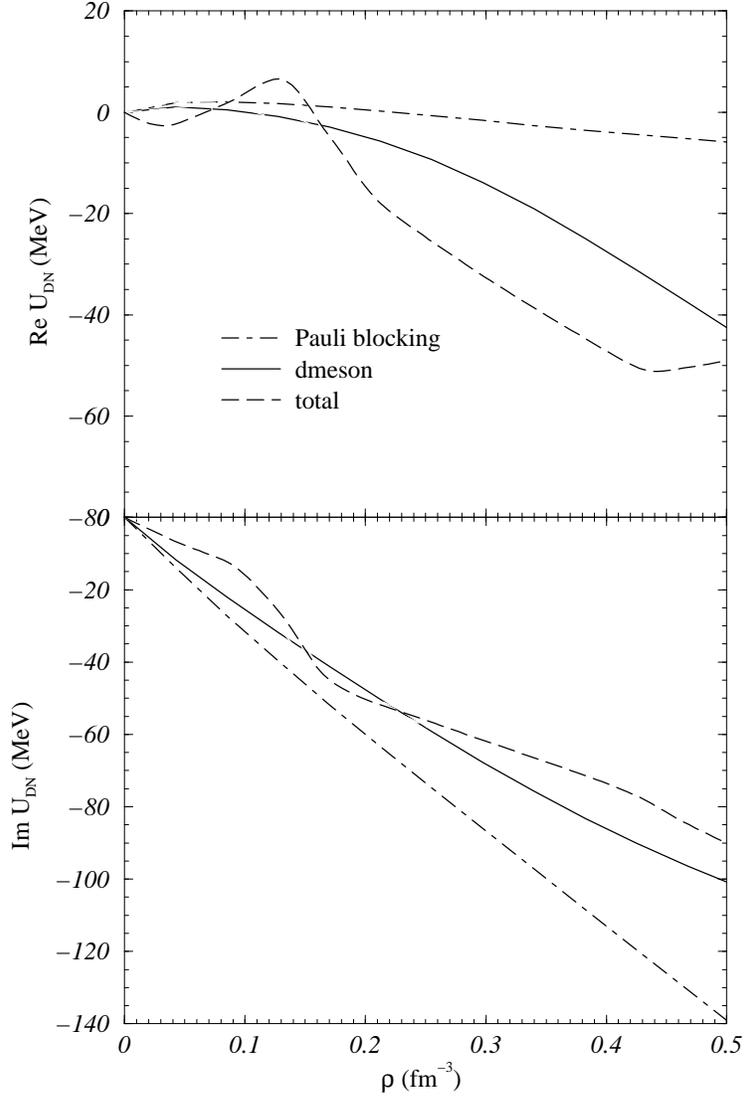}
}
      \caption{\small Real and imaginary parts of the $D$-meson  potential at $k_D=0$ as a function of density for  $\Lambda=1$ GeV and $g^2=$13.4  and for different approaches: in-medium calculation including only Pauli blocking (dot-dashed lines), self-consistent calculation for the $D$-meson (solid lines) and self-consistent calculation for the $D$-meson including the dressing of nucleons and the pion self-energy (long-dashed lines).
}
        \label{fig:dmeson4}
\end{figure}

In Fig.~\ref{fig:dmeson4} we plot the real and imaginary parts of the $D$-meson  potential at $k_D=0$ as a function of density for  $\Lambda=1$ GeV and $g^2=$13.4 and for the three previous in-medium calculations: in-medium calculation with only Pauli blocking (dot-dashed lines), self-consistent calculation for the $D$-meson (solid lines) and self-consistent calculation for the $D$-meson including the dressing of nucleons and the self-energy of pions (long-dashed lines). The $D$-meson potential becomes more attractive as we increase the density in all three approaches, showing a less smooth behaviour in the case when nucleons and pions are dressed both in the real and imaginary parts. However, the  potential turns out to be more attractive for a self-consistent calculation with respect to the case when only Pauli blocking is considered. If we compare these results to the case of the  $\bar K$  potential, the picture depicted here is  different. For the $\bar K$ meson case, the shift of the $\Lambda(1405)$ resonance in energy due to the Pauli blocking changes the $\bar K N$ scattering amplitude at the threshold from being repulsive to attractive giving rise to an attractive $\bar K$  potential. However, when the properties of $\bar K$ were incorporated self-consistently, the attraction was drastically reduced \cite{Tolos01,Ram00,Lutz98a}. For the $D$-meson, the $D N$ threshold is 2.806 GeV and, therefore, the $D N$ amplitude is studied for energies  on the right-hand side of the free $\Lambda_c(2593)$ resonance, away from this resonant structure. Furthermore, in the $\bar K$ meson case, the s-wave $I=0$ $\bar K N$ amplitude turns out to be the main contribution to the $\bar K$  potential due to the presence of the $\Lambda(1405)$ resonance. For the $D$-meson, as the energy region  of interest is not sitting on the resonance, the $I=1$ could also become important in the calculation of the $D$-meson  potential. 

\begin{figure}[htb]
\centerline{
     \includegraphics[width=0.6\textwidth,angle=-90]{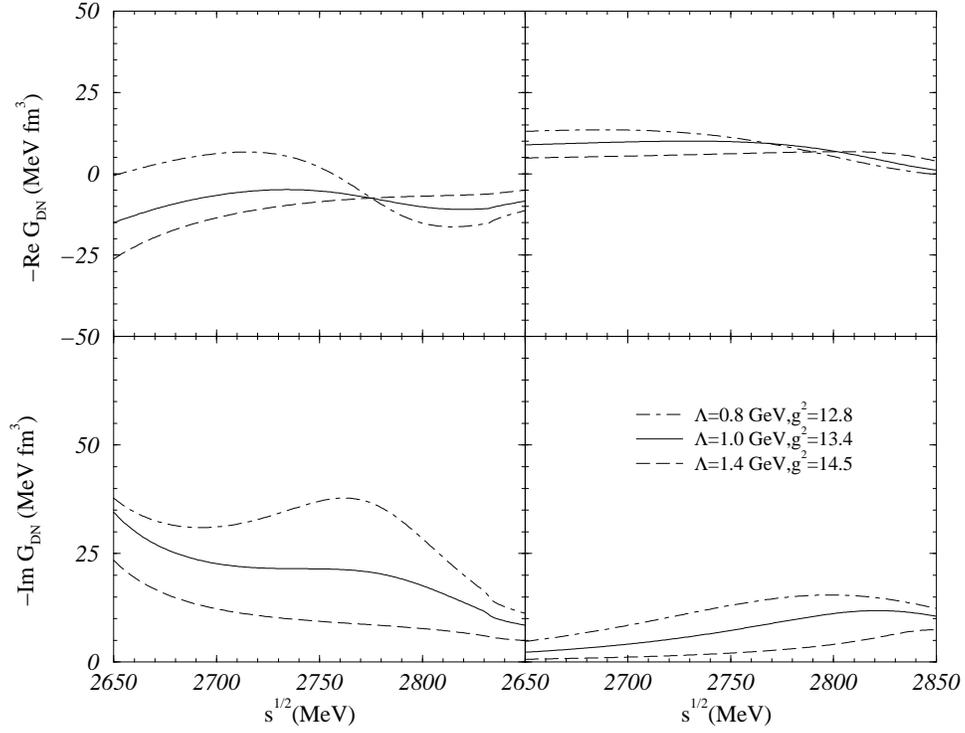}
}
      \caption{\small  Real and imaginary parts of the s-wave $DN$ amplitude at $\rho=\rho_0$ in the $I=0$ (left panels) and $I=1$ (right panels) channels as functions of the center-of-mass energy  at total momentum $\mid\vec{k}_D+\vec{k}_N\mid=0$ for the self-consistent calculation for the $D$-meson and for different sets of coupling constants and cutoffs.
}
        \label{fig:dmeson5}
\end{figure}

\begin{figure}[htb]
\centerline{
     \includegraphics[width=0.6\textwidth,angle=-90]{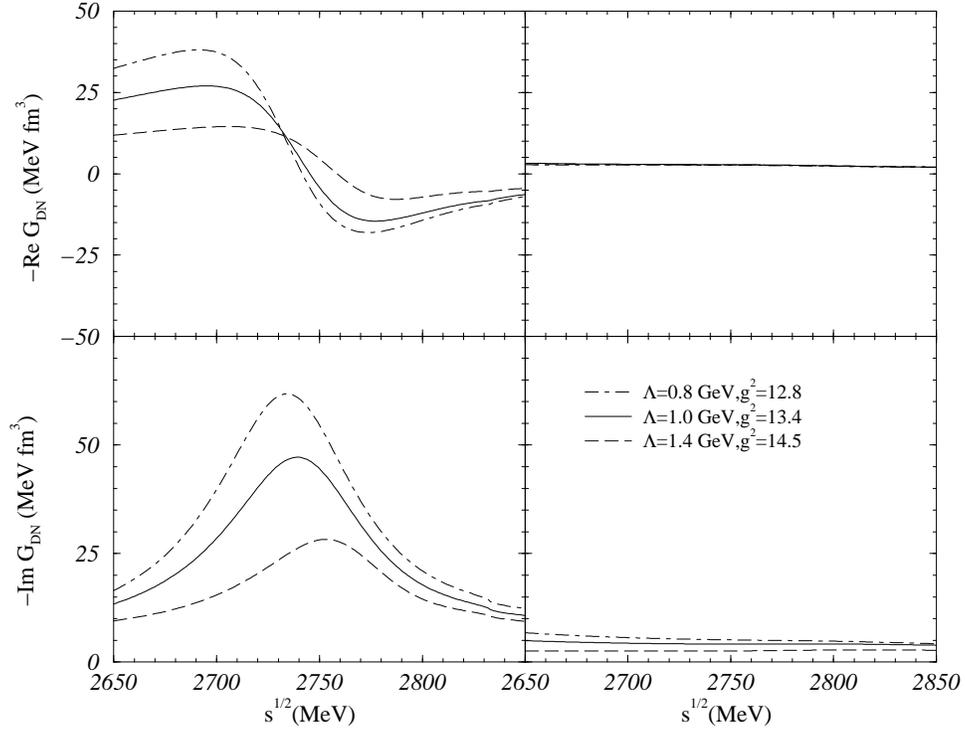}
}
      \caption{\small Real and imaginary parts of the $DN$ amplitude at $\rho=\rho_0$ in the $I=0$  (left panels) and $I=1$  (right panels) channels as functions of the center-of-mass energy  at total momentum $\mid\vec{k}_D+\vec{k}_N\mid=0$ for the self-consistent calculation for the $D$-meson including the dressing of nucleons and the pion self-energy and for different sets of coupling constants and cutoffs.
}
        \label{fig:dmeson6}
\end{figure}

\begin{figure}[htb]
\centerline{
     \includegraphics[width=0.6\textwidth,angle=-90]{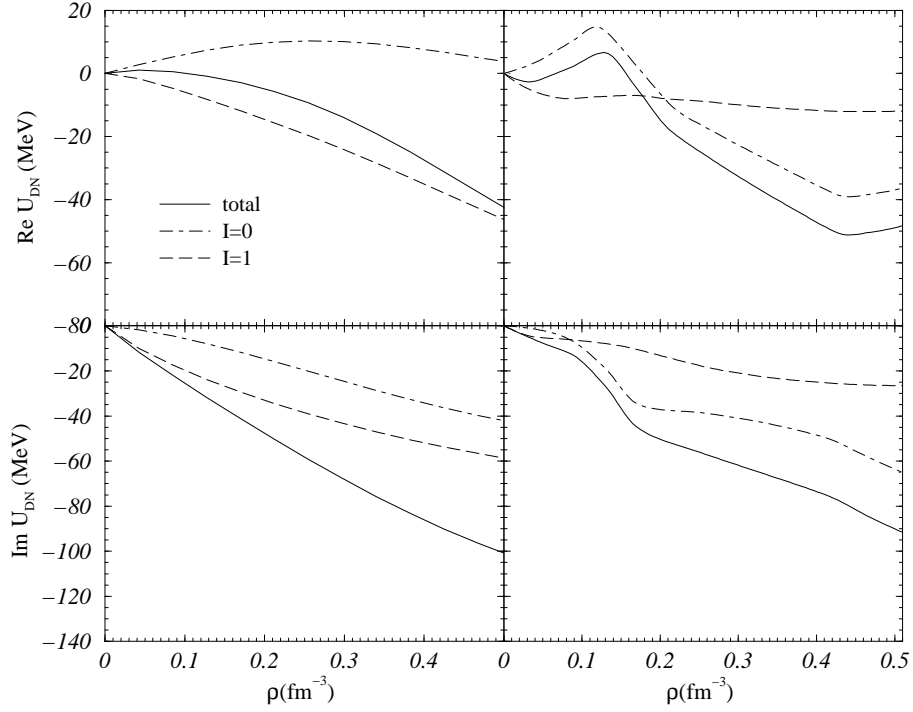}
}
      \caption{\small  Real and imaginary parts of the $D$-meson  potential at $k_D=0$ as a function of the density for  $\Lambda=1$ GeV and $g^2=$13.4 including the isospin decomposition for two approaches: self-consistent calculation for the $D$-meson (left panels) and self-consistent calculation for the $D$-meson including the dressing of nucleons and the pion self-energy (right panels). 
}
        \label{fig:dmeson7}
\end{figure}

In order to study the isospin dependence of in-medium $D N$ interaction in more detail as well as the dependence on the chosen set of coupling constants and cutoffs, in Fig.~\ref{fig:dmeson5} and \ref{fig:dmeson6} we represent the
real and imaginary parts of the $DN$ amplitude at $\rho=\rho_0$ in the $I=0$  and $I=1$  channels as functions of the center-of-mass energy around the $D N$ threshold  at total momentum $\mid\vec{k}_D+\vec{k}_N\mid=0$ for the self-consistent calculation when only the $D$-meson is dressed (Fig.~\ref{fig:dmeson5}) and for the full self-consistent calculation (Fig.~\ref{fig:dmeson6}). Furthermore, in Fig.~\ref{fig:dmeson7} we show the 
 real and imaginary parts of the $D$-meson potential at $k_D=0$ as a function of the density for  $\Lambda=1$ GeV and $g^2=$13.4 including the  isospin decomposition.

For the first self-consistent approach where only $D$-mesons are dressed at $\rho=\rho_0$ (Fig.~\ref{fig:dmeson5}), the real parts of the $I=0$ and $I=1$ components turn out to be of the same order of magnitude. Therefore, it is not clear which isospin will determine the behaviour of the $D$-meson  potential. Actually, the  potential  depends on the exact value of both contributions taking into account the isospin factor $2I+1$ for each isospin component (see Eq.~(\ref{eq:upot1})).
For $\Lambda=1$ GeV, the main contribution for the potential at $\rho=\rho_0$ comes from the $I=1$ component, as  seen in Fig.~\ref{fig:dmeson7}. On the other hand, we note that for $\Lambda=0.8$ GeV, the $I=0$ component moves from attraction to repulsion as the energy increases (Fig.~\ref{fig:dmeson5}). Nevertheless, the region of integration to obtain the $D$-meson potential for all the sets of parameters of coupling constants and cutoffs lies around 2.80-2.84 GeV at $\rho=\rho_0$ and, therefore, we obtain a repulsive effect for  $I=0$ and an attractive effect for $I=1$ in the  potential at $\rho=\rho_0$ for all sets (see Fig.~\ref{fig:dmeson7} for $\Lambda=1$ GeV). The imaginary part displays a smoother behaviour than in the case of the full self-consistent calculation. When the full calculation is performed (Fig.~\ref{fig:dmeson6}), the peak observed around energies of 2.75 GeV  gets diluted with increasing cutoff. However, the real part shows more structure than in the previous self-consistent approach and the $I=0$ component becomes more important with respect to the previous case. This effect together with a less important contribution of the $I=1$ component makes the $I=0$ govern the $D$-meson potential for densities larger than $\rho=1.5\rho_0$ for all sets. For $\rho=\rho_0$ and $\Lambda=1$ GeV, the attractive $I=1$ component still dominates. What it is  not so clear is whether the $I=0$ will be attractive or repulsive for all the sets as the region of integration lies around 2.71-2.78, where the $I=0$ real part changes in sign. The different region of the integration in this last case with respect to the case where only the $D$-meson is dressed is mainly due to the inclusion of an attractive potential for the nucleons as the value of the $D$-meson  potential does not change by a great amount for both self-consistent processes (see Fig.~\ref{fig:dmeson7}).

In Fig.~\ref{fig:dmeson7} we observe that, in the case where only the $D$-meson is dressed and for $\Lambda=1$ GeV, the real part of the $D$-meson potential is governed by the $I=1$ component as  density grows. This is also seen for the other set of parameters studied. On the other hand, for the full calculation, the $I=0$ component controls the behaviour of the $D$-meson potential as we increase the density not only for $\Lambda=1$ GeV but also for the other sets. This is due to the resonant structure seen in Fig.~\ref{fig:dmeson6} that also causes a less smooth behaviour of the $D$-meson  potential.

\begin{figure}[htb]
\centerline{
     \includegraphics[width=0.6\textwidth,angle=-90]{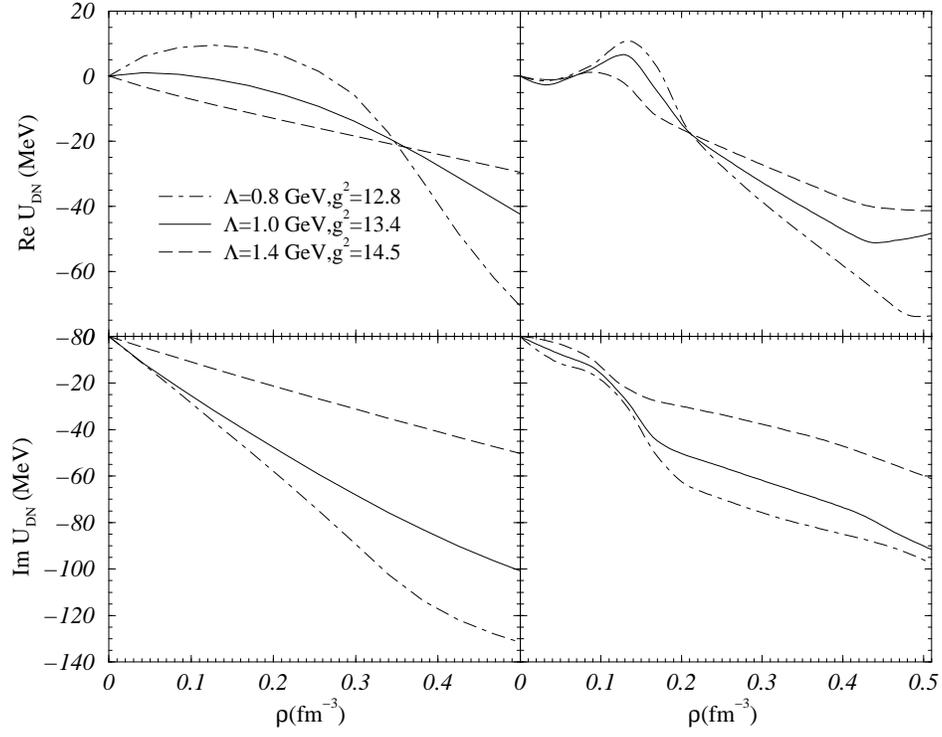}
}
      \caption{\small Real and imaginary parts of the $D$-meson  potential at $k_D=0$ as a function of the density for different sets of coupling constants and cutoffs and for two approaches: self-consistent calculation for the $D$-meson (left panels) and self-consistent calculation for the $D$-meson including the dressing of nucleons and the pion self-energy (right panels).
}
        \label{fig:dmeson8}
\end{figure}

In order to study the dependence on the cutoff and coupling constant, 
we show in Fig.~\ref{fig:dmeson8} the $D$-meson  potential 
at $k_D=0$ as a function of the density for different sets of coupling constants and cutoffs and for the two approaches mentioned before. 
When only the $D$-meson is dressed, higher densities are required in order to obtain an attractive potential with respect to the full self-consistent calculation. In both cases, the $D$-meson  potential tends to get more attractive as density increases. The reason is that,  in order to obtain the $D$-meson potential as density increases, we are integrating over a larger energy region where the interaction is attractive. 
However, as mentioned in the previous plot, the isospin component responsible for that behaviour is different depending on the approach. On the other hand, the value of the potential turns out to be slightly sensitive to the chosen set of parameters. There is, however, in both cases a density for which all sets give the same value for the $D$-meson potential. When only the $D$-meson is dressed,  we obtain a 
range of values for the $D$-meson potential at $\rho=\rho_0$ between  8.6 MeV for $\Lambda=0.8$ GeV and -11.2 MeV for $\Lambda=1.4$ GeV. For the full self-consistent calculation, the range of values covered lies in between 2.6 MeV for $\Lambda=0.8$ GeV  and -12.3  MeV for $\Lambda=1.4$ GeV. 
This result would indicate that our self-consistent coupled-channel calculation gives
a different prediction for the $D$-meson  potential as compared to the calculations based on the QCD sum-rule approach \cite{arata,weise}, the quark-meson coupling model \cite{qmc}, lattice calculations \cite{digal,wong,lattice} and a recent work  based on a chiral model that obtains 
the medium modification
of the mass of D-meson due to its interaction
with the light hadron sector \cite{dmeson}. The calculations based on the
QCD sum rule approach as well as quark-meson coupling model predict
the mass drop of the D-meson to be of the order of 50-60 MeV
at nuclear matter density. A similar drop at finite temperatures
\cite{digal} is suggested from the lattice calculations 
for heavy quark potentials. An interaction Lagrangian based
on chiral perturbation theory also gives a similar shift
at nuclear matter density,
when the Tomozawa-Weinberg interaction is supplemented
by the attractive DN Sigma term and the charm content
of the nucleon is ignored \cite{dmeson}. On the other hand, the effective 
chiral Lagrangian approach  \cite{dmeson} is seen to yield a larger drop
of the D-meson masses ($\simeq$ 200 MeV at nuclear matter density).
In the present calculation, the coupled-channel effects seem to result in an overall drastic reduction of the in-medium effects independent of the parameters ($g$,$\Lambda$) and the in-medium properties of the intermediate states compared to the previous works.

\begin{figure}[htb]
\centerline{
     \includegraphics[width=0.6\textwidth,angle=-90]{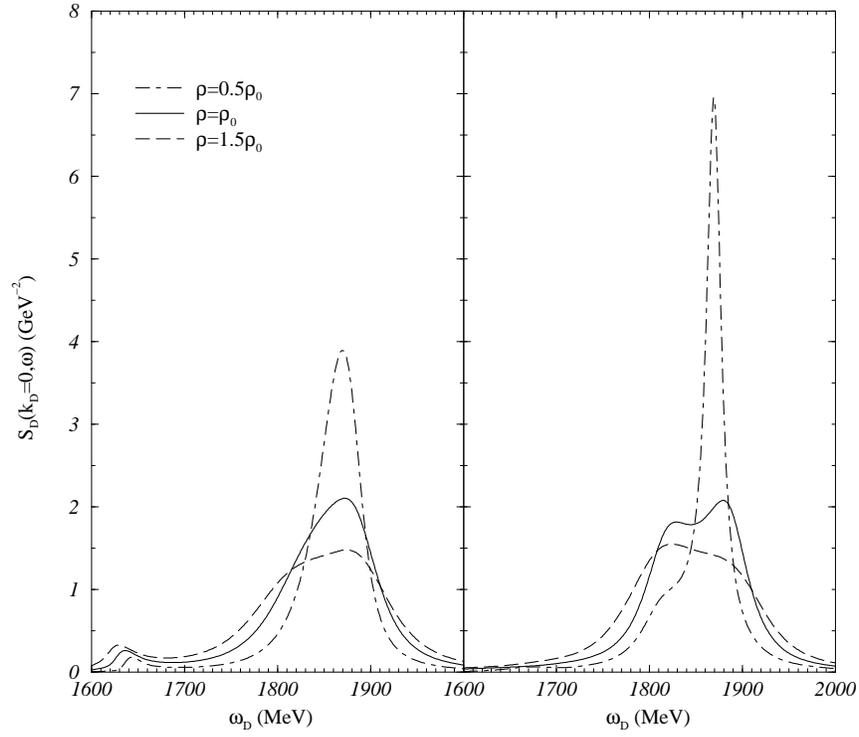}
}
      \caption{\small $D$-meson spectral density at $k_D=0$ as a function of energy with  $\Lambda=1$ GeV and $g^2=$13.4 for different densities and for two approaches: self-consistent calculation for the $D$-meson (left panels) and self-consistent calculation for the $D$-meson including the dressing of nucleons and the pion self-energy (right panels).
}
        \label{fig:dmeson9}
\end{figure}

Finally, once self-consistency is reached, we calculate the full energy dependence of the $D$-meson self-energy which defines the in-medium $D$-meson single-particle propagator and its spectral density (see Eqs.~(\ref{eq:prop}-\ref{eq:spec})). The spectral density at zero momentum is shown in Fig.~\ref{fig:dmeson9} for $\Lambda=1$ GeV and for several densities in the  two approaches considered before. In the case when only the $D$-meson is dressed in the self-consistent process, the peak of the $D$-meson pole moves towards lower energies as density increases since the $D$-meson potential becomes more attractive for $\Lambda=1$ GeV (see Fig.~\ref{fig:dmeson8}). Moreover, it is observed that the $D$-meson spectral density falls off more slowly on the left-hand side of the quasiparticle peak as density increases. This is due to the $I=1$ component of the off-shell $D$-meson self-energy, which is related to the $D$-meson potential through Eq.~(\ref{eq:relation}). As density increases, we have already discussed that the $I=1$ component governs the behaviour of the  potential and, hence, the self-energy. Therefore, any structure lying close the quasiparticle peak is enhanced and clearly visible in the spectral density. On the other hand,
we also notice some structure of the spectral density to the left of the quasiparticle peak at energies of the $D$-meson of around 1.63-1.65 GeV, the origin of which could be traced back to the presence of the $\Lambda_c(2593)$ resonance. 

For the full self-consistent calculation, the quasiparticle peak moves towards lower energies as density increases according to the behaviour of the $D$-meson  potential seen in Fig.~\ref{fig:dmeson8}, although there is region of densities around $0.5\rho_0$ where the $D$-meson potential turns out to be repulsive. For densities between $\rho_0$ and 1.5 $\rho_0$, the quasiparticle peak mixes with a structure, already noticed on the left-hand side of the quasiparticle peak for 0.5  $\rho_0$, making it more difficult to separate the contribution of the quasiparticle peak and the effect of this structure. This bump in the spectral density is due to the structure observed  in the imaginary part of the $I=0$ component of the in-medium $D N$ interaction which lies close to the $D N$ threshold. The nature of this structure and in-medium changes of the D-meson production could probably be studied experimentally in the near future 
with the PANDA experiment at the GSI International Facility in Darmstadt \cite{ritman}.

\section{Conclusions}
\label{sec:conclusions}

We have performed a microscopic self-consistent coupled-channel calculation of the single-particle potential of the $D$-meson and, hence the $D$-meson spectral density, embedded in symmetric nuclear matter assuming a separable potential for the s-wave $DN$ interaction. 

The $\Lambda_c(2593)$ resonance, which is the counterpart of the $\Lambda(1405)$ in the charm sector, has been generated dynamically for a given set of coupling constants $g$ and cutoffs $\Lambda$. 

We have also studied the medium effects on the $\Lambda_c(2593)$ resonance and, hence on the $D$-meson potential, due to the Pauli blocking of the intermediate nucleonic states as well as due to the dressing of nucleons and pions. The $D$-meson optical potential has been  obtained for different approaches: in-medium calculation including only Pauli blocking on the intermediate nucleonic states, self-consistent calculation for the $D$-meson and self-consistent calculation for the $D$-meson including the dressing of nucleons and the pion self-energy. 

We observe that the two  self-consistent schemes show a stronger density dependence compared to the case when only Pauli blocking effects are included. When only the $D$-meson is dressed in the self-consistent procedure, 
the $D$-meson potential at $\rho=\rho_0$ stays between  8.6 MeV for $\Lambda=0.8$ GeV and -11.2 MeV for $\Lambda=1.4$ GeV. For the full self-consistent calculation, the range of values covered lies in between 2.6 MeV for $\Lambda=0.8$ GeV and -12.3 MeV for $\Lambda=1.4$ GeV. We conclude that the coupled-channel effects seem to result in an overall reduction of the in-medium effects independent of the set of parameters ($g$,$\Lambda$) and the in-medium properties of the intermediate states compared to  previous work. 

The isospin dependence of the $D$-meson in both approaches have been also analyzed. While the $I=1$ amplitude governs the behaviour of the $D$-meson potential when only $D$-mesons are dressed, the $D$-meson is basically determined by the $I=0$ component in the full self-consistent calculation.

The $D$-meson spectral density has been finally obtained for both self-consistent schemes. Although the quasiparticle peak stays closer to its free position in both cases for nuclear matter saturation density, the features of the low-energy region on the left-hand side of the quasiparticle peak are different according to the different in-medium behaviour of the $\Lambda_c(2593)$ resonance in both approaches. 

The in-medium effects devised in this work can be studied in heavy-ion
experiments at the future International Facility at GSI. The PANDA
experiment at GSI \cite{ritman} will measure hadrons with charm by
antiproton beams on nuclei with its microvertex detector.  In-medium
changes of open charm hadrons can be addressed by the study of the
excitation function and the correlation function of $D$- and $\bar D$-mesons.
We stress that,
although the in-medium potential for $D$-mesons has turned out to be quite small
in our investigation, the production of $D$-meson in the nuclear medium
will be still enhanced due to the additional strength of the $D$-meson
spectral function below the quasiparticlepeak. This effect is similar to the one
extracted for the enhanced $\bar K$ production in heavy-ion collisions
\cite{Tolos_effect}. 

The present coupled-channel approach to the $D$-meson properties in the
nuclear medium is, as the first of its kind, exploratory and can be
improved by incorporating relativistic potentials as well as chiral
constraints on the bare hadronic interactions. It will be also
interesting to explore the in-medium effects for $D$-mesons in dynamical
approaches for studying e.g.\ the excitation function, which we will
leave for future work.

\section*{Acknowledgments}

The authors are very grateful to A. Ramos for critical reading of the manuscript. L.T. wishes to acknowledge the financial support from the Alexander von Humboldt Foundation. A.M. is grateful to the Institut f\"ur Theoretische Physik for
warm hospitality and acknowledges financial support from
Bundesministerium f\"ur Bildung and Forschung (BMBF).

\newpage


\end{document}